

\tolerance = 30000
\baselineskip=16pt

\def\cmm2{{\,\rm cm^{-2}}}
\def\cm2{{\,{\rm cm}^2}}
\def\cmm3{{\,{\rm cm}^{-3}}}
\def\gcmm3{{\,{\rm g\,cm^{-3}}}}

\def\ga{\mathrel{\mathpalette\fun >}}
\def\fun#1#2{\lower3.6pt\vbox{\baselineskip0pt\lineskip.9pt
  \ialign{$\mathsurround=0pt#1\hfil##\hfil$\crcr#2\crcr\sim\crcr}}}

\rightline{FERMILAB-Pub-92/381-A}
\rightline{December 1992}
\rightline{submitted to {\it Physics Letters B}}

\bigskip
\bigskip
\bigskip

\centerline {\bf CONSTRAINTS ON DIRAC NEUTRINOS FROM SN 1987A}
\bigskip
\medskip
\bigskip
\centerline {\bf R. Mayle$^{\rm a}$, D.N. Schramm$^{\rm b,c}$,
Michael S. Turner$^{\rm b,c}$, and J.R. Wilson$^{\rm a}$}

\vskip 0.5in
\centerline{$^{\rm a}$Lawrence Livermore National Laboratory}
\centerline{Box 808}
\centerline{Livermore, CA~~94550~~~USA}
\medskip
\centerline{$^{\rm b}$Departments of Physics and Astronomy \&
Astrophysics}
\centerline{Enrico Fermi Institute}
\centerline{The University of Chicago}
\centerline{5640 S. Ellis Avenue}
\centerline{Chicago, Illinois~~60637-1433~~~USA}
\medskip
\centerline{$^{\rm c}$NASA/Fermilab Astrophysics Center}
\centerline{Fermi National Accelerator Laboratory}
\centerline{Box 500}
\centerline{Batavia, Illinois~~60510-0500~~~USA}

\vskip 1 in
\centerline{\bf Abstract}
\bigskip
\noindent The Livermore Supernova Explosion Code was used to calculate
the effect of a massive Dirac neutrino on neutrino
emission from SN 1987A in a fully self-consistent manner.  Spin-flip
interactions lead to the copious emission of sterile, right-handed
neutrinos and cool the core faster than the observed neutrino
emission time for Dirac masses exceeding about 3$\,$keV.
This limit is relaxed to $7\,$keV if pion emission processes
in the core are neglected. These limits are compared with
the previous less stringent limits of Burrows et al.

\vfill\eject

\beginsection{Introduction}

Many authors [1] noted that spin-flip reactions enable
the weakly interacting left-handed components of Dirac neutrinos
to change to sterile right-handed neutrinos and thereby
free-stream out of the central core of a nascent neutron star.
This free streaming, rather than the usual
diffusion [2], leads to a rapid cooling of the core and thereby
shortens the duration of the neutrino burst.
Since the SN1987A neutrino burst
was observed by the Kamiokande (KII) [3] and the Irvine-Michigan-Brookhaven
(IMB) [4] detectors to last of order
$10\,$sec, the KII and IMB results can be used
to constrain Dirac neutrino
masses.  The production of sterile right-handed
neutrinos is proportional to the
neutrino mass squared, and so the larger the mass, the more rapid the
cooling.  Based upon numerical cooling models that
incorporated the cooling effects of right-handed
neutrinos, Gandhi and Burrows
set a mass limit of 28 keV [1].  With the interest generated by the
possible existence of a 17 keV
neutrino, this limit was re-examined [5].
This re-examination culminated in a calculation that
included the neutrino-nucleon spin-flip scattering,
nucleon-nucleon bremsstrahlung,
and pion emission processes, allowed for the effects of degeneracy, and
explored a range of cooling models [6].  The result was a very
conservative mass limit of $25\,$keV
and a limit of $15\,$keV when pions are assumed to
be as abundant as nucleons, with an
estimated uncertainty of $10\,$keV.

Given that this limit is indecisively close
to 17 keV and, in general, to
help clarify the sensitivities to the cooling model, we decided to carry
out a similar calculation with the Livermore Supernova Explosion
Code [2].  The Livermore Supernova Explosion Code
has a somewhat higher central
temperature than the models used by Burrows et al. [6].  The
Livermore Code follows the
pre-collapse evolution of a massive star [7] that fits the
characteristics of the progenitor to SN 1987A, which leads to a
collapsing core that is on a higher temperature
adiabat.  In addition, the equations of state used in the Livermore
Code are somewhat different from those of Burrows et al. [6],
particularly with regard to the treatment of pions.  For all
these reasons, we felt it important to explore the limits that
follow from the Livermore Code. The emphasis of this paper will be
those aspects of the
calculation that differ from Burrows et al. [6].
While the final fate of the 17 keV neutrino obviously lies in
the hands of the experimentalists, it is nonetheless interesting
to see what constraints the SN 1987A neutrino observations
imply.  We note that there are other stringent limits
based upon SN 1987A that apply to unstable neutrinos [8,9].

\beginsection{The Model}

The general model of gravitational collapse as it relates to
neutrino emission and to SN 1987A is reviewed in Ref. [10].
The Livermore Supernova Explosion Code is described in detail in
Refs. [2, 11, 12].  As discussed in Ref. [12], the code has
been ``tuned'' to fit the observed aspects of SN 1987A,
including the pre-collapse progenitor,
neutrino emission, and the explosion energy.
In addition, the present calculations are based upon an improved
equation of state, which yields a significantly higher meson
density in the core, but does not otherwise alter the
agreement with SN 1987A.  The high pion (and Kaon)
densities follow from modelling
heavy-ion collisions at the Bevalac with a mean-field equation
of state of the type used in stellar-collapse calculations [13].
High meson densities are a consequence of
finite-density effects on the energy-momentum
relation for pions (and Kaons), and receive theoretical support from
recent work by Politzer and Wise [14] and
others [13, 15], indicating that Kaon-condensates are to be expected in
nuclear matter.  In addition, the work of McAbee and Wilson [13]
indicates that Bevalac heavy-ion experiments
support these ideas.  While these
equation-of-state effects are of small importance to the
basic collapse observables, they have a dramatic effect on the
production of sterile, right-handed neutrinos due to
pion-nucleon scattering production of right-handed neutrinos,
$$\pi + N \rightarrow N + \nu_R \bar \nu_R .\eqno(1)$$
The rates we have used for right-handed neutrino production
are the same as those used by
Burrows et al. [6], and all three processes they considered are considered
here. Thus, the key difference is the equation of state
(and hence the densities of pions and other mesons and the
temperatures).

We use the pion mean-field model [13].  This
model has the following equations for the energy-momentum
dispersion relation:

$$\epsilon^2 = m^2_\pi + p^2 \left [ 1 + {\Lambda^2}\chi \over {1
-g^\prime \Lambda^2 \chi} \right ] ; \eqno(2)$$

$$\Lambda^2 \chi = {{-4.52 \omega \rho \exp [-(p/7m_\pi)^2]} \over
{m^2_\pi (\omega^2 - \epsilon^2)}} ; \eqno(3)$$

$$\omega (p) = \sqrt{m_\Delta ^2 + p^2} -m_N ; \eqno(4)$$

\noindent where $m_\pi$, $m_N$, $m_\Delta$ are respectively
the pion, nucleon and delta masses, $g^\prime$ is the Landau-Migdal
parameter, and $\rho$ is the nucleon density.  By reproducing
the experimental results, both pion numbers and spectra from Bevalac
experiments involving La on La
collisions at 1.35, 0.74, 0.53 GeV/N [13], McAbee
and Wilson determined a fit for
the Landau-Migdal parameter, $g^ \prime = 0.50 + 0.06
\rho/\rho_n$, where $\rho_n$ is nuclear matter
density, 0.16 fm$^{-3}$.  The same equation of state
used in these heavy-ion collision calculations was
used in the present supernova calculations. The temperature and
densities spanned in the Bevalac experiments are comparable to those
encountered in the regions of the proto-neutron star
where right-handed neutrino production is the largest.

To incorporate the pions into the supernova calculations, we
determine the pion abundances, $n_{\pi^+}$,
$n_{\pi^-}$, and  $n_{\pi^0}$, by requiring
chemical equilibrium, $\mu_{\pi^-}= - \mu_{\pi^+} = (\mu_n -
\mu_p)$, and charge balance, $n_{e^+} + n_p + n_{\pi^+} - n_{\pi^-}
- n_{e^-} = 0$.  In addition, $\beta$-equilibrium implies that
$\mu_e - \mu_{\nu_e} = \mu_n -\mu_p$.
The high electron chemical potential encountered in the central regions of
the core of a proto-neutron star without pions is greatly
reduced by the presence of negative pions, and the ``degeneracy
energy''  released increases the temperature considerably
(see Fig. 1).

In our calculations, right-handed neutrino
losses are implemented as follows.  Energy loss due to
neutrino-nucleon spin-flip scattering
($\nu_L + N \rightarrow \nu_R + N$) is given by:
$$\dot\epsilon_{SF} = {{\mu_{\nu_e} G^2_F m^2_\nu (m_N T) ^{3/2}}
\over {2^{9/2} \pi^5}} \lbrack 1.2 I_s (r_p) + 1.4 I_s
(r_n) \rbrack ; \eqno(5)$$
where the approximation for $I_s$ used is given by
$$I_s (r)^{-1} =  \ {{2e^{- r}} \over {\sqrt{\pi}}} + {
1 \over {\sqrt{1 + \mid r \mid}}} - { 1 \over {8(1 + \mid r
\mid)^{3/2}}} ; \eqno(6)$$
where $r_{p, n} = (\mu_{p, n} - m_N) / T.$  This
energy loss, is apportioned to the matter
energy and the neutrino energy in the ratio of 2:1.  Thus, every third
spin flip caused a loss in lepton number.

Energy loss due to nucleon-nucleon
bremsstrahlung processes ($N + N \rightarrow N + N + \nu_L \bar \nu_L$)
is given by:
$$\dot\epsilon_{BR} = {{160f^4 g^2_A G^2_F m^2_\nu m^{9/2}_N
T^{13/2}} \over {15 \pi^2 m^4_\pi}} \lbrack 0.5 (I_b (r_p, r_p) + I_b (r_n,
r_n)) +3 I_b (r_p, r_n) \rbrack ;$$
$$I_b (r_1, r_2) = 2.39 \times 10^5 \left[ e^{-r_1 -r_2} +  0.25
(e^{-r_1} +e^{-r_2}) \right] +$$
$$\left\{ {{1.73 \times 10^4} \over {(1 + \mid{{r_1 + r_2}
\over 2} \mid)^{1/2}}} + {{6.92 \times 10^4} \over {(1 +
\mid {{r_1 + r_2} \over 2} \mid)^{3/2}}} + {{1.73
\times 10^4} \over {(1 + \mid {{r_1 + r_2} \over 2}
\mid)^{5/2}}} \right\} ; \eqno(7)$$
\noindent where $f \sim 1$ is the pion-nucleon coupling and $g_A
\sim 1.25$ is the axial-vector coupling.

Energy loss due to pion processes ($\pi + N \rightarrow N + \nu_L \bar \nu_L$)
is given by:
$$\dot\epsilon_\pi = {{2^{10} g^2_A f^2 G^2_F m^2_\nu} \over {
105 \pi^{7/2} m^2_\pi}} T^3 n_N n_\pi ; \eqno(8)$$
with $n_\pi = n_{\pi^-} + n_{\pi^+} + n_{\pi^0}$.  The losses
from bremsstrahlung and pion processes are taken out of the
matter energy only. In all our calculations we take the tau
neutrino to be the massive neutrino (so far as supernova physics
goes, it could just as well be the muon neutrino as
the tau and mu neutrinos are treated identically). We also assume
that all three neutrino species are in chemical equilibrium with $\mu_\nu =
\mu_e$.  The rates used here are described in detail in Ref. [17].

In the present calculations we have ignored the presence of
Kaons.  If we apply the  Kaon mass reduction formula
of Refs. [14, 15, 18] to the conditions in the central part
of the proto-neutron star at the time of large right-handed
neutrino losses, the presence of Kaons would lead to the
additional release of electron-degeneracy energy, about 20 to 30 MeV
per nucleon.  This energy release would further increase
the temperature.  We have not included the
Kaons in our calculations for energy loss because the
parameters involved are not well established at present, and thus we are
probably underestimating right-handed neutrino losses since
they increase rapidly with temperature.

In general, our high pion densities yield higher core temperatures which
leads to less degenerate conditions, thereby exaggerating the differences
between Burrows' treatment and the Livermore Code.
Since the Bevalac data implies a
high pion density [13], and since the Livermore
Code's initial conditions are well fit to initial models of the
SN 1987A progenitor [7], we feel that the more stringent constraints we
obtain here should be taken seriously.
We also note that merely adding pions, as was
done in Ref. [6], cannot mimic all the pion effects we have
found here, because the presence of pions
also leads to higher core temperatures.

We note, in passing, that high pion/Kaon densities should also
affect axion production.  Since these pion/Kaon effects were not
included in previous SN 1987A
axion mass limits [16], we intend to explore the high pion/Kaon
density effects on axion constraints in a subsequent paper. A
preliminary estimate suggests that the mass limit of $10^{-3}$eV
may improve to $10^{-4}$eV [17].

Because of the immense time it takes to run models
where the feedback of right-handed neutrino emission
on the cooling of the
core is taken into account, we carried out some calculations
without full feedback to explore sensitivity to parameter
choices.  These runs also serve to illustrate how important feedback
is in determining constraints.
Since much discussion in the literature has focused on the
effect of electron, neutrino, and nucleon degeneracy on right-handed neutrino
emission [5,6], it is worth noting that in our models
all degeneracy effects are properly taken into account.
Moreover, temperatures are sufficiently high that
all species are non-degenerate to a good
approximation, making the point moot.

\beginsection{Results}

Figure 1 shows the core temperature 0.5 seconds after core
bounce.  As discussed above, the equation of state
with high pion density results in significantly higher core temperatures
than that without low pion density.  As time goes on, the peak temperature
(at ${\rm \sim 0.5 M_\odot}$ in Fig.~1) moves inward.  It
is worth noting that the density and absolute value of the
neutrino chemical potential is also higher in the central core
when pions are included, but $\mu_\nu$/T is lower.

Figure 2 shows the time evolution of the radius of the neutrino
``photosphere'' for a 10 keV Dirac neutrino compared with a zero
mass neutrino.  Time is measured from core bounce.
Only at late times is the difference significant.

Figure 3 shows the average energy
of the proper-helicity anti-electron neutrinos (the neutrinos
responsible for the KII and IMB events) versus time
for zero mass and a mass of 10 keV; as expected, this energy
is insensitive to the Dirac neutrino mass.

Figure 4a shows the luminosity in 10 keV
right-handed neutrinos for all three emission processes in a
calculation with full feedback.  Note that for a 10 keV
neutrino, the neutrino-nucleon spin-flip process is large
and cools the inner core so fast that the pion processes barely dominant,
and then only at late times.
Thus, even without the pion processes, a 10 keV
mass Dirac neutrino would lead to unacceptably rapid cooling.
Figure 4b is the same as 4a but for a mass of 3.16 keV.
Here the pion processes are more significant. Figure 4c shows the
right-handed neutrino luminosities for a 1 keV Dirac neutrino
with \underbar{no} feedback.  Note that the pion processes are about
100 times larger than the other two. Since the right-handed
neutrino emission rates vary as
$m^2_\nu$, it is clear that pion processes would dominate and
rapidly cool the core for masses as small as a fraction of a keV
when feedback is neglected, and hence a reliable estimate of the
mass limit requires taking feedback into account.

Figure 5 shows a comparison of the proper-helicity anti-electron
neutrino luminosity in full feedback models with pion processes
for $m_\nu = 0$, 3.16 and 10 keV. For a mass greater than 3 keV or
so, the duration of the $\bar \nu_e$ burst is significantly
shortened, and thus a Dirac neutrino mass larger than
this can be excluded by the KII and IMB results.

\beginsection{Conclusions}

In summary, based upon our calculations employing full feedback and the
Livermore Gravitational Supernova Explosion Code,
and incorporating an equation of state that reproduces the
Bevalac heavy-ion experiments, we use the duration of the
neutrino bursts observed by KII and IMB to exclude a Dirac
neutrino mass greater than about 3 keV.
Including the presence of Kaons in
the post-collapse core  would only strengthen our limit.  Even if
pion emission processes are ignored, it still
appears difficult to accommodate a mass greater than 10 keV with the
Livermore Code.  The difference between the Burrows et al. [6]
``very conservative'' limit of 25 keV and the 10 keV limit
obtained (here ignoring the pion processes) traces to the higher
core temperatures in our models. Further, the present results
indicate that the Burrows et al. limits were, as stated, very
conservative, and suggest that the KII and IMB data very
probably exclude a 17 keV neutrino.
\bigskip
\bigskip

\beginsection{Acknowledgements}

We thank George Fuller for his early involvement in this work.
We are also grateful to Stan Woosley for providing the model
for the progenitor of SN 1987A used in our calculations.
We acknowledge many useful discussions with members of ``the
17 keV neutrino experimental community'' who encouraged us to complete this
study.  We also thank Adam Burrows and Raj Gandhi for valuable
discussions about their work, and Georg Raffelt, David
Seckel and Rocky Kolb for their comments.
This work was supported in part by the DOE and NSF at Livermore, by NSF,
DOE (Nuclear and HEP) and NASA
at the University of Chicago, and by the DOE and NASA (through
grant NAGW 1321) at Fermilab.

\bigskip

\beginsection{References}

\item{[1]} R. Gandhi and A. Burrows, Phys. Lett. B {\bf 246}, 19
(1990); {\it ibid} {\bf 261}, 519 (1991);
G.G. Raffelt and D. Seckel, Phys. Rev. Lett. {\bf 60},
1793 (1988);  K. Gaemers et al., Phys. Rev. D {\bf 40}, 309 (1989);
J.A. Grifols and E. Masso, Nucl. Phys. {\bf B 331}, 244
(1990); {\it ibid.} Phys. Lett. B {\bf 242}, 77 (1990);
A. Perez and R. Gandhi, Phys. Rev. D {\bf 41}, 2374 (1990).

\item{[2]} See e.g., R. Mayle, J.R. Wilson, and D.N. Schramm, ApJ {\bf 318},
288 (1987).

\item{[3]} K. Hirata et al., Phys. Rev. Lett. {\bf 58}, 1490 (1987).

\item{[4]} R.M. Bionta et al., Phys. Rev. Lett. {\bf 58}, 1494 (1987).

\item{[5]} M.S. Turner, Phys. Rev. D {\bf 45}, 1066 (1992);
 R. Gandhi and A. Burrows, Phys. Lett. B {\bf 261}, 519
(1991);  G.G. Raffelt and D. Seckel, unpublished (1992).

\item{[6]} A. Burrows, R. Gandhi, and M.S. Turner, Phys. Rev.
Lett. {\bf 68}, 3834 (1992); the cooling models used in these
calculations are discussed by A. Burrows and J. Lattimer, ApJ
{\bf 309}, 178 (1986) and A. Burrows, ApJ {\bf 334}, 891 (1988).

\item{[7]} S. Woosley and T. Weaver, Phys. Rep. {\bf 163}, 79 (1988).

\item{[8]} S. Dodelson, J.A. Frieman, and M.S. Turner, Phys. Rev.
Lett. {\bf 68}, 2572 (1992).

\item{[9]} E.W. Kolb and M.S. Turner, Phys. Rev. Lett. {\bf 62},
509 (1989).

\item{[10]} D.N. Schramm and J.W. Truran, Phys. Rep. {\bf 189},
89 (1990).

\item{[11]} R. Mayle, Ph.D. Thesis, University of California (1987).

\item{[12]} R. Mayle and J.R. Wilson, Livermore Report UCRL-97355 (1987).

\item{[13]} T. McAbee and J.R. Wilson, Livermore preprint (1992);
R. Mayle, M. Tavani, and J.R. Wilson, Livermore preprint (1992).

\item{[14]} H.D. Politzer and M. Wise, Phys. Lett. B {\bf 273},
156 (1991);  D. Montano, H.D. Politzer, and M. Wise, Nucl. Phys. {\bf
B 375}, 507 (1992).

\item{[15]} G.E. Brown, K. Kubodeva, and M. Rho, K.K. \& G.E.B.
preprint \#6, Stony Brook (1992).

\item{[16]} see e.g., M. Burrows, M.S. Turner, and R.P. Brinkman, Phys.
Rev. D {\bf 39}, 1020 (1989);
R. Mayle, J.R. Wilson, J. Ellis, K. Olive, D.N. Schramm,
and G. Steigman, Phys. Lett. B {\bf 219}, 515 (1989).

\item{[17]} M.S. Turner, Phys. Rev. D {\bf 45}, 1066 (1992).

\item{[18]} B. Friedman, V.R. Panharipande, and Q.N. Usmani,
Nucl. Phys. A {\bf 372}, 483 (1981).

\vfill\eject
\beginsection{Figure Captions}

\bigskip

\noindent Fig. 1.  The core temperature as a function of interior
mass 0.5 seconds after core bounce.  Note the large effect
that the high density of pions has on the core temperature.

\bigskip
\noindent Fig. 2.  The time evolution of the radius of the neutrino
``photosphere'' for Dirac masses of 0 and 10 keV.

\bigskip
\noindent Fig. 3. The average energy of the proper-helicity $\bar \nu_e$'s
as a function of time after core bounce for Dirac masses of 0
and 10 keV.

\bigskip
\noindent Fig. 4. The energy loss rates in right-handed neutrinos as a
function of time after core bounce for the three emission
processes: (a) $m_\nu = 10$ keV with feedback; (b) $m_\nu =
3.16$ keV with feedback; (c) $m_\nu = 1$ keV with no feedback.

\bigskip
\noindent Fig. 5. The luminosity in proper-helicity $\bar \nu_e$s for
$m_\nu$ = 0, 3.16, 10 keV with full feedback and pion processes
included.  Note that for $m_\nu \ga {\cal O}($3 keV) the
duration of the neutrino burst is significantly shortened, and
thus a Dirac mass this large can be excluded by the KII and IMB data.

\end